\documentclass[%
 aip,
 amsmath,amssymb,
 reprint,%
]{revtex4-1}
\pdfoutput=1
\usepackage{graphicx}
\usepackage{dcolumn}
\usepackage{bm}

\usepackage[utf8]{inputenc}
\usepackage[T1]{fontenc}
\usepackage{mathptmx}
\usepackage{etoolbox}

\usepackage{enumitem}

\makeatletter
\def\@email#1#2{%
 \endgroup
 \patchcmd{\titleblock@produce}
  {\frontmatter@RRAPformat}
  {\frontmatter@RRAPformat{\produce@RRAP{*#1\href{mailto:#2}{#2}}}\frontmatter@RRAPformat}
  {}{}
}%
\makeatother
\begin{document}

\preprint{AIP/123-QED}

\title[Limitations to the energy resolution of single-photon sensitive Microwave Kinetic Inductance Detectors]{Limitations to the energy resolution of single-photon sensitive Microwave Kinetic Inductance Detectors}
\author{M. De Lucia}
\affiliation{Dublin Institute for Advanced Studies, 31 Fitzwilliam Place, D02XF86, Dublin, Ireland}
\affiliation{Trinity College Dublin, School of Physics, College Green, Dublin, Ireland}
\email{delucia@cp.dias.ie}
\author{G. Ulbricht}
\affiliation{Dublin Institute for Advanced Studies, 31 Fitzwilliam Place, D02XF86, Dublin, Ireland}
\affiliation{Maynooth University, Physics Department, Maynooth, Co. Kildare, Ireland}
\email{ulbrichtg@cp.dias.ie}
\author{E. Baldwin}
\affiliation{Dublin Institute for Advanced Studies, 31 Fitzwilliam Place, D02XF86, Dublin, Ireland}
\affiliation{Trinity College Dublin, School of Physics, College Green, Dublin, Ireland}
\author{J. D. Piercy}
\affiliation{Dublin Institute for Advanced Studies, 31 Fitzwilliam Place, D02XF86, Dublin, Ireland}
\affiliation{Trinity College Dublin, School of Physics, College Green, Dublin, Ireland}
\author{O. Creaner}
\affiliation{Dublin City University, School of Physical Sciences, Glasnevin Campus, Dublin, Ireland}
\affiliation{Dublin Institute for Advanced Studies, 31 Fitzwilliam Place, D02XF86, Dublin, Ireland}
\author{C. Bracken}
\affiliation{Maynooth University, Physics Department, Maynooth, Co. Kildare, Ireland}
\affiliation{Dublin Institute for Advanced Studies, 31 Fitzwilliam Place, D02XF86, Dublin, Ireland}
\author{T. P. Ray}
\affiliation{Dublin Institute for Advanced Studies, 31 Fitzwilliam Place, D02XF86, Dublin, Ireland}
\affiliation{Trinity College Dublin, School of Physics, College Green, Dublin, Ireland}
\date{\today}

\begin{abstract}
This paper describes the energy resolution of Microwave Kinetic Inductance Detectors (MKIDs), and models some limiting factors to it. Energy resolution is a measure of the smallest possible difference in energy of the impinging photons, $\Delta E$, that the detector can identify and is of critical importance for many applications.\\
 Limits to the energy resolution cause by the Fano effect, amplifier noise, current inhomogeneities, and readout sampling frequency are  taken into consideration for this model. This paper describes an approach to combine all of these limitations and predict a wavelength dependency of the upper limit to the resolving power.
\end{abstract}

\maketitle
\begin{figure}[h!]
    \centering
    \includegraphics[height=0.82\textheight]{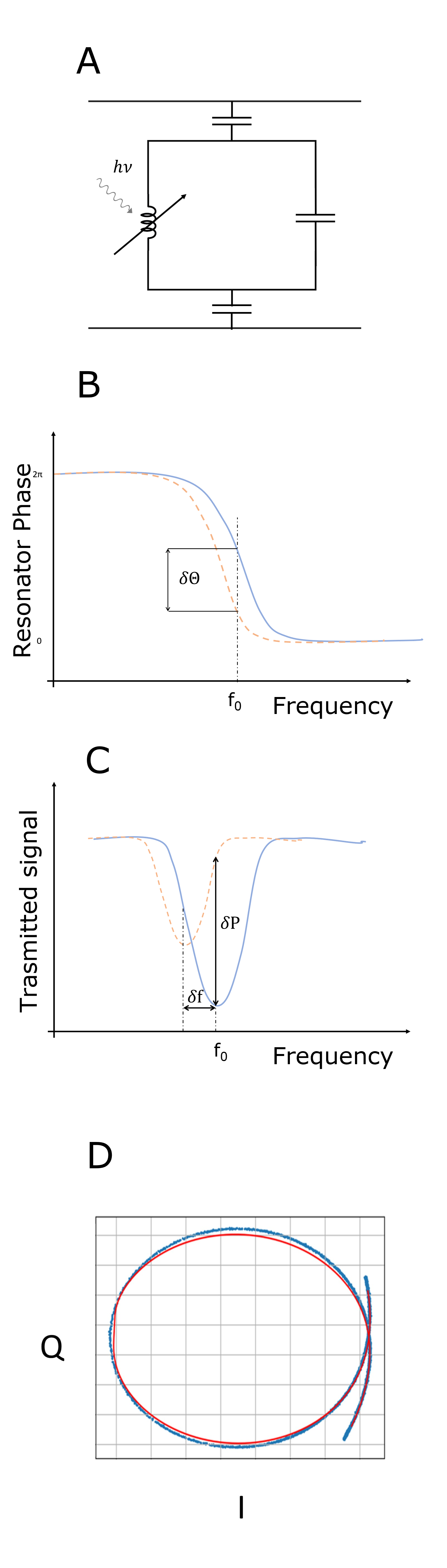}
    \caption{(A) schematic of a lumped-element MKID: an L-C circuit capacitively coupled to a microwave feedline and to ground. The variable inductor represents the photosensitive element of the detector and the change in kinetic inductance as consequence of a detection event. (B) dip in transmission power at the resonant frequency of an MKID. (C) Phase change around the resonance frequency of an MKID. In orange: the phase shift induced by a detection event.  (D) IQ plane frequency sweep of an MKID plotted in the IQ plane.}
    \label{fig:kids_wp}
\end{figure}

Microwave Kinetic Inductance Detectors  (MKIDs) are cryogenic photo-detectors and can be used in a broad range of astronomical applications from X-Rays to the far-infrared \cite{mkidsapplications}.
In the Ultra Violet, Optical and Near-Infrared (UVOIR) part of the spectrum MKIDs can achieve single photon sensitivity and are capable to distinguish the energy of individually detected photons. In this regime,  their design is based on lumped-element superconducting L-C resonators \cite{Day2003,lekids} and their working principle is shown in Figure \ref{fig:kids_wp}.  Upon striking the superconductor, the photon is absorbed and, as it breaks  a number of Cooper-pairs\cite{Day2003}, it produces a cascade of unpaired electrons\cite{deVissersphononiccrystal} (called quasi-particles) and phonons. This sudden reduction in superconducting carrier density can be seen as a change in inductance of the L-C circuit\cite{Pippard1,Mazin,Zmuidzinas}, and therefore as a variation of its resonance frequency. A much more thorough description can be found in e.g. \citet{Day2003} and \citet{Zmuidzinas}. This change in resonance frequency can be monitored as a change in amplitude or phase\cite{eoin-thesis} of the complex transmission as shown in Figure \ref{fig:kids_wp}. With a time constant $\tau_{qp}$, that is material dependent \cite{qp1,qp2}, the quasi-particles  will recombine into Cooper-pairs and the detector returns to its idle state ready for a new event.  \\
In reality, an MKID response signal is better monitored through a combination of coordinates in the complex IQ plane\cite{eoin-thesis}. A frequency sweep across an MKID in the IQ plane is described by a circle\cite{gao,eoin-thesis}, and is shown in Figure \ref{fig:kids_wp} (d). A detection event in the IQ plane is described by a trajectory along this circle proportional to the phase change in the complex transmission. \\
Details on the geometry, the design and the electromagnetic simulations of the MKIDs taken into consideration for this paper can be found in the supplementary material.\\
The line-width of the resonator, and the diameter of the circle in the IQ plane, are linked to the total quality factor of the resonator $Q_{tot} \cite{gao}$ which is defined  as:
\begin{equation}
    \frac{1}{Q_{tot}}= \frac{1}{Q_i} + \frac{1}{Q_c}
\end{equation}
where $Q_c$ is a measure of the coupling of the L-C resonator to the feedline and $Q_i$ accounts for all others losses \cite{Zmuidzinas,gao}. 
\\
\newline
The main figure of merit for UVOIR MKIDs is energy resolution $\Delta E$.  An ideal detector, when illuminated by monochromatic light, produces a spectrum that is represented by an infinitely sharp line \cite{WRLeo}.  In reality, the width of any spectral line is finite. The energy resolution of a detector is defined as $\Delta E$, , the full-width-half-maximum of the linewidth; correspondigly  $\frac{E}{\Delta E}$ is known as the  Resolving Power R. 
\\
At present,
\citet{deVissersphononiccrystal} demonstrate that the best resolving power achieved for an MKID corresponds to  $\frac{E}{\Delta E} \sim 55$ at 402 nm. We will discuss, in order, the following effects and how they induce theoretical upper limits to the resolving power: \\
\begin{enumerate}
    \item Statistical processes in the MKID. \\These are well known and we will just present a
  		short summary from literature to provide a more complete picture

    \item Amplifier noise and how its standard nomenclature of ‘noise temperature’ can be
 		translated into resolving power limitations.

    \item Current density inhomogeneity
    \item Sampling frequency and their influence on R depending on readout details.
\end{enumerate}{}
\vspace{0.3cm}
First, we give an overview of the statistical effects induced in the MKID during the detection event. When a  photon strikes on the superconductor, it generates a single high-energy phonon which will cascade rapidly into broken Cooper pairs and lower energy phonons\cite{Day2003, EnergyDownconversion}. As soon as these secondary phonons have less than twice the energy of the superconducting bandgap, they cannot break further Cooper pairs and the energy converted in such phonons is therefore lost to the detection principle. The amount of broken Cooper pairs shows statistical fluctuations, and Guo et al.\cite{Guo} describe the best resolving power that an otherwise perfect MKID can achieve as a pair-breaking detector is given by $$ \frac{E}{\Delta E} = 0.425\sqrt{\frac{\eta h\nu}{F\Delta_s}}$$ where $\eta\approx 0.57$ is a typical value for the pair-breaking efficiency \cite{EnergyDownconversion}, $h\nu$ is the energy of the impinging photon, and $\Delta_s = 1.764 k_B T_c$ is the superconducting gap energy of the adsorbing material. $F$ is the Fano factor, which describes all statistical processes that do not lead to pair-breaking. For MKIDs, a typical value for the Fano factor\cite{Mazin}  is $F=0.2$  These values yield a theoretical maximum of $ R =\frac{E}{\Delta E}$ of $114.5$ for $\lambda_0 = 400$ nm photons and a $T_c$ of $0.8$ K. While the value of F is not expected to vary significantly, other kinds of energy resolving detectors, such as gas and silicon based ionisation detectors, exhibit a Fano factor that varies quite significantly with the material \cite{Fano_material,fano_gas1}.  \\

Next to the Fano effect, the factor that limits the resolving power of photon-counting MKIDs most is the intrinsic noise of High Electron Mobility Transistor (HEMT) amplifiers. We will use the Low Noise Factory LNF-LNC4$\_$8C as an example. It operates at 4K and is well suited for the 4-8 GHz band.
Its nominal noise temperature18 is given as 2.1 K by the manufacturer. HEMT-amplifier add random noise to the signal of the MKID before amplification. In the IQ plane we assume that the I and the Q components are equally affected. A photon detection by an MKID causes the measured signal
in the IQ plane to move along a semi-circle overlapping with the frequency-sweep circle (see Fig \ref{fig:kids_wp} D). This movement is usually expressed as a change of the so-called phase angle be-
tween the measured IQ spot and the centre of the frequency sweep loop. The HEMT-amplifier noise is added to this phase signal, therefore the height of the phase jump generated by
photon detection has an additional uncertainty given by the noise contribution of the HEMT-amplifier. As both internal and coupling quality factors of the MKID resonator influence
the size of the frequency-sweep loop they have to be taken into account; for an in-depth description of the formula used to generate Fig. \ref{fig:HEMT_QQ}, and Fig. 2 please see the supplementary material .
We will refer to the uncertainty of the phase-pulse as $\Delta \alpha$ where $\alpha$ is the height of the phase pulse. The ratio $\frac{\alpha}{\Delta\alpha}$ is roughly equivalent to $\frac{E}{\Delta E}$ and therefore to the resolving power R.\\
We are also assuming an MKID optimised for an $\alpha = \pi$ phase shift when struck by a  photon of wavelength $\lambda_0 = 400$ nm. 

\begin{figure}[ht!]
\includegraphics[width=0.45\textwidth]{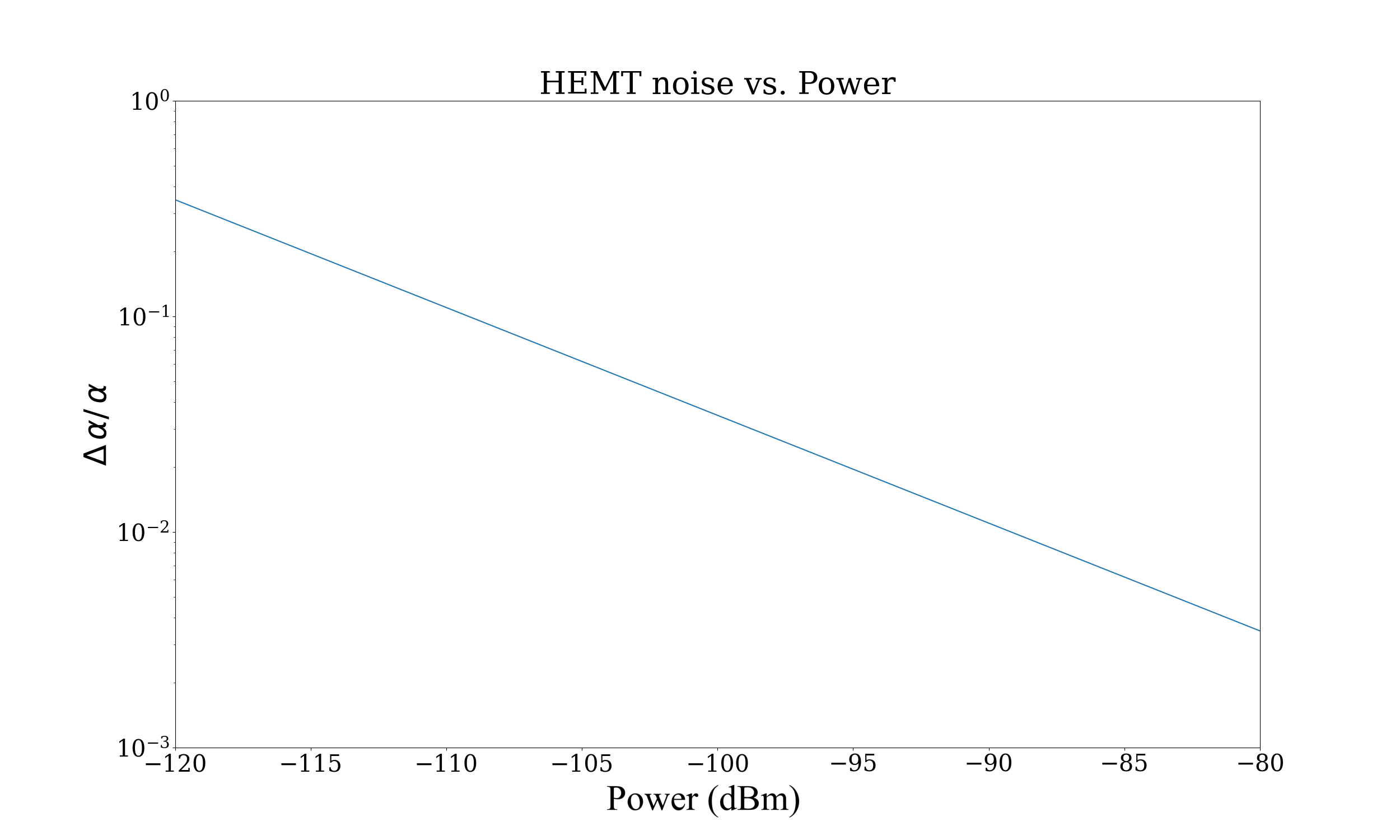}
  \label{fig:HEMT_power} 
  \caption{HEMT noise, $\frac{\Delta\alpha}{\alpha}$ as a function of the power at the resonator. assuming $Q_i = 10^5$, $Q_c = 3\times10^4$ and a noise temperature of 2.1 K. Typical operating powers for UVOIR MKIDs are in the -90 dBm range.}
\end{figure}{}
As amplifier noise depends inversely on input power (See Appendix 1.B) it is more pronounced the lower the signal to be amplified is. MKIDs are driven at significantly varying power levels depending on superconductor, pixel geometry, here we therefore just assume a generated power (Pg, see appendix) at the location of the MKID and, assuming no further losses within the MKID, of the HEMT-amplifier of about $-90$ dBm, realistic for our setup. With further assumptions of $Q_i =10^5$, $Q_c=3\times10^4$ and the above mentioned  HEMT-amplifier noise, we get $\approx 91.1$. This corresponds to a noise fluctuation of the measured phase angle of$\pm 1.98^{\circ}$  which is consistent with the commonly observed $\pm5 ^{\circ}$ noise in the baseline. \\
\begin{figure}[h!]
    \includegraphics[width=0.45\textwidth]{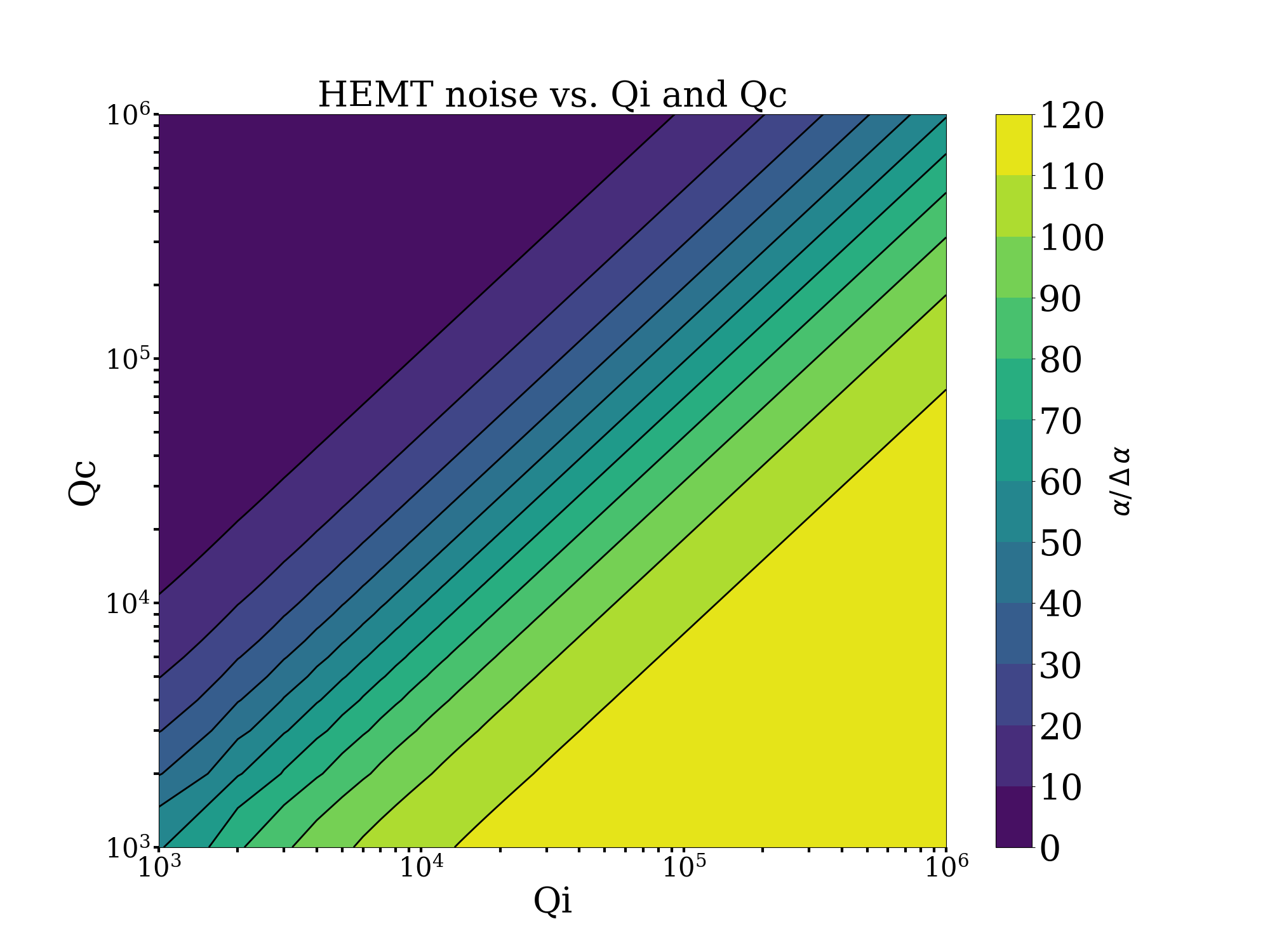}
    \caption{2D graph showing the dependency of the limit to R casued by the HEMT-amplifier noise  in degrees, assuming that a photon produces a $180^\circ$ signal, as a function of the coupling quality factor $Q_c$ and the internal quality factor $Q_i$.}
    \label{fig:HEMT_QQ}
\end{figure}{}
\newline
 Figure 2 shows the noise contribution due to the HEMT amplifier as a function of the Pg under the assumption of an internal quality factor $Q_i = 10^5$ and coupling quality factor $Q_c = 3\times 10^4$.
Figure \ref{fig:HEMT_QQ} shows a 2-Dimensional plot of R limited by the HEMT-amplifier noise as depending on both the quality factor $Q_i$ and the coupling quality factor $Q_c$. The plot shown refers to a power of $-90$ dBm as seen by the MKID.
\vspace{0.5cm}

Inhomogeneity of the current density along the inductor can be another source of uncertainty in the response of an MKID and could therefore limit the achievable R. If different parts of the superconducting inductor exhibit different current densities the signal resulting from otherwise-identical absorbed photons will depend on their strike positions, as the change in kinetic inductance will be different because of kinetic inductance non-linearities \cite{kher,Zmuidzinas}.
Current density uniformity depends on the resonator's geometry: e.g. current crowding at right angles is a well known effect for both normal metals and superconductors \cite{current-crowding}. Furthermore, in a lumped element MKID charge carriers slow down towards both ends of the inductor, where it connects to the capacitor. Our simulations suggest that this effect can reach up to a $5\%$ change in current density for our example geometry. All the current inhomogeneity effects discussed here are of course highly dependent on the chosen MKID geometry. We show the results for the example we have chosen, but the steps necessary to achieve these results (see supplementary material) will be more interesting for general use.\\

To fully evaluate the effects of current density inhomogeneities on resolving power we simulated current density with SONNET \cite{sonnet} with a grid size  of $1\mu{m}$  and imported these into a Python  script  developed by De Lucia and available upon request.  For a figure please see the supplementary materials. Our working assumption is that each element of the inductor, as defined by the simulation's through the grid size, contributes to the total kinetic inductance of the resonator equally and the total kinetic inductance is given by the the sum of these contributions. Our example resonator has a resonant frequency is $f_0 = 6$ GHz and $Q_c = 30\,000$. 

We assume a well optimised resonator at the impinging wavelength, and thus impose  a depletion of Cooper-pairs that produces a phase shift  $\alpha = 0.95\,\pi$ if the photon strikes where the current is most uniform. We also simplified by ignoring quasi-particle dynamics, assuming that a photon hitting one of these inductor elements only changes the kinetic inductance in this element. Our script then translates current density into expected MKID signal height $\alpha$, for details please see the supplementary materials.\\
\begin{figure}[h!]
\centering
    \includegraphics[width=0.45\textwidth]{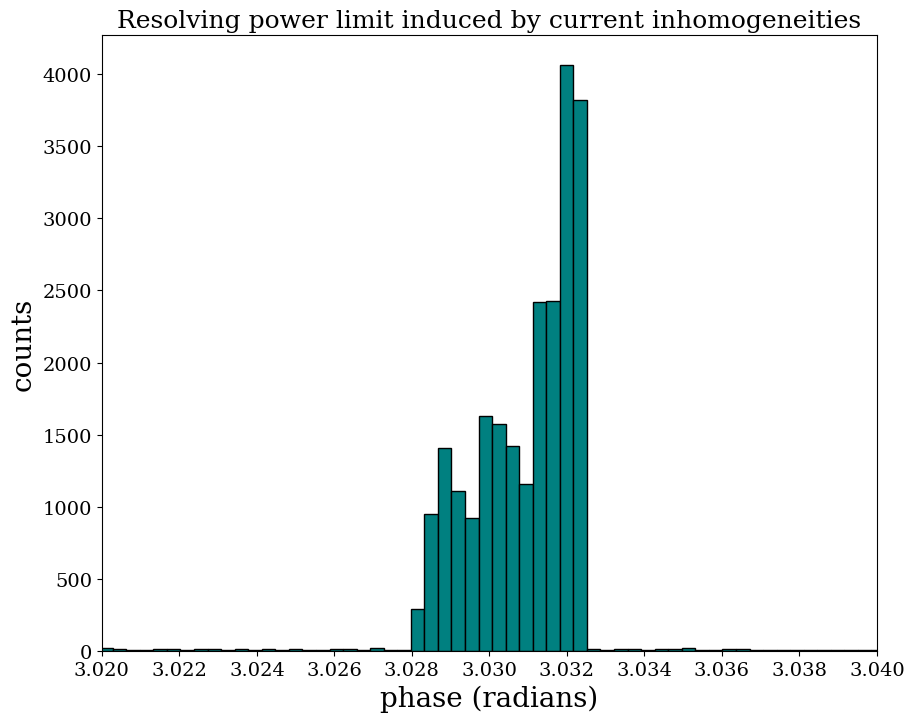}
  \label{fig: hist} 
  \centering
  \includegraphics[width=0.5\textwidth]{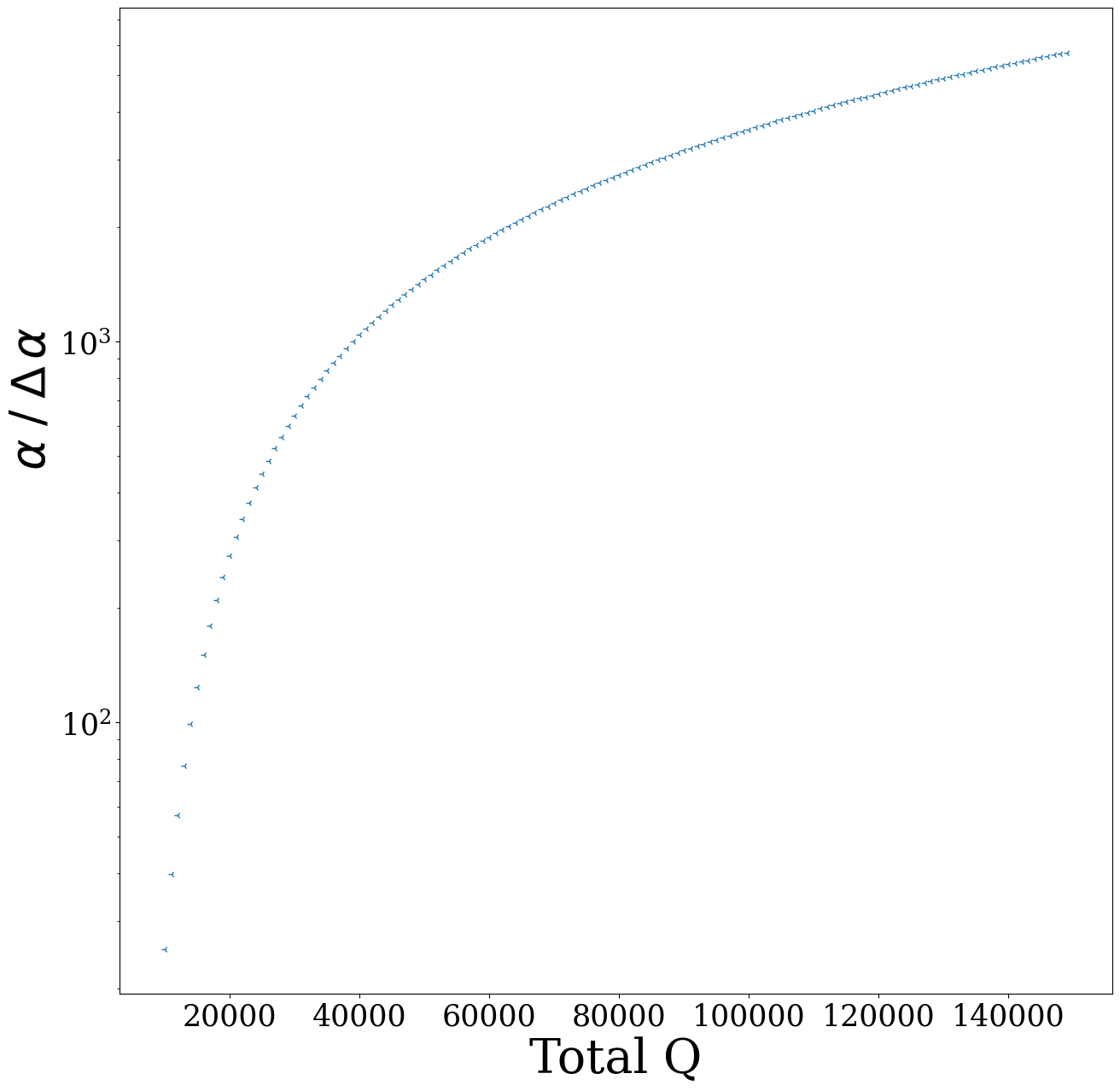}
  \caption{Top: Broadening of the energy resolution of an  MKID  due to simulated current inhomogeneities. The width of the distribution represents the energy resolution of a resonator whose designed resonant frequency is $6$ GHz and the total quality factor is $Q_{tot} = 30000$.  Bottom: Dependence of R as a function of $Q_ {tot}$}
  \label{fig:ci}
\end{figure}{}
As previously discussed, the Energy Resolution $\Delta E$, can be described in terms of $\frac{\Delta \alpha}{\alpha}$. Figure \ref{fig:ci} (Top)  shows the broadening of the energy resolution of an MKID with non-homogeneous current density assuming every inductor element is hit by the same photon. Given the chosen grid-size, the current density simulations resulted in three main values that represent the three main peaks in Figure \ref{fig:ci} (Top). The three peak structure is caused by the chosen grid-size, as this resulted in a corresponding current density simulation. The width of this profile represents  $\Delta \alpha$, Based on this simulated example MKID, our simulated current inhomogeneities would limit resoling power to R = 606

This value depends on the total quality factor $Q_{tot}$ of the resonator as shown in Figure \ref{fig:ci} (Bottom). Our model suggests that an improvement can be obtained by increasing the resonator's total quality factor, but Figure \ref{fig:HEMT_QQ} shows that increasing the total quality factor increases the HEMT-amplifier noise and the current inhomogeneity has an in comparison a much lower impact. Our model does not include dissipative effects induced by the excited quasi-particles, which behave like normal-state electrons. These losses account for a reduction in the internal quality factor $Q_i$ and have not been included in the discussion. \\

\begin{figure}[h!]
    \centering
    \includegraphics[width=0.45\textwidth]{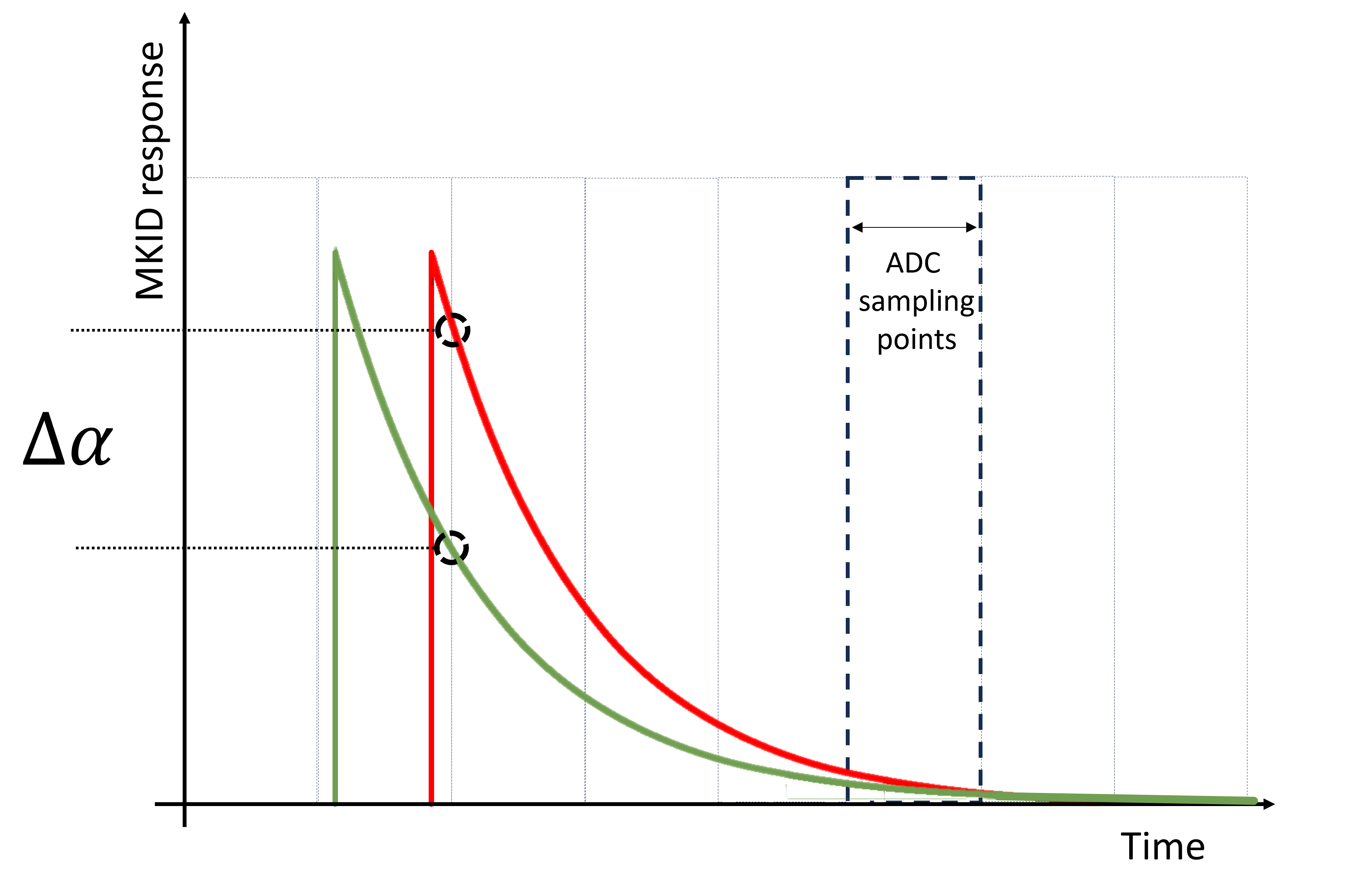}
    \caption{Schematic of the exponential pulse decay (ignoring all other noise sources) for two identical photons hitting the detector just before (red) or just after (green) an ADC sample. The signal decays with the quasi-particle recombination time $\tau_{qp}$, and the ADC is sampling at constant time intervals $\Delta t$. The induced variation $\Delta \alpha$ represents the upper limit for the energy resolution caused by the ADC sampling speed.}
    \label{fig:sampling}
\end{figure}

The last effect we want to discuss  that could limit the resolving power of UVOIR MKIDs is the interplay between the ADC sampling rate  during digital measurement and the quasi-particle lifetime $\tau_{qp}$ of the superconductor of choice. As different readout schemes are popular, we will discuss the single-pixel, homodyne readout first: Here the MKID signal is down-mixed to DC in one step and I and Q channels are directly monitored with ADCs, often with a sampling rate of 1 MHz. In this case the effects of the digital sampling rate on R need to be taken into account if the sampling time interval starts to no longer be significantly below $\tau_{qp}$ (as shown in Fig. \ref{fig:sampling}): 

When a photon hits an MKID, broken Cooper pairs will start to recombine immediately, resulting in an exponential decay of the signal with $\tau_{qp}$ as decay time. As most commonly used ADCs only take snapshots of a varying input voltage\cite{dataconverters}, they are not able to distinguish between a higher energy photon that has been absorbed just after a measured data point and a less energetic one hitting the MKID just before a measurement point (see Fig. \ref{fig:sampling}): If a photon hits an MKID just before an ADC sample it generates its maximum possible signal $\alpha$. But if the same photon would have been absorbed just after the preceding ADC sample, the signal would have the whole sampling time to decay. As photons arrive at uncontrolled times this effect results in an inaccuracy $\Delta \alpha$ of the measured signal equivalent to the maximum signal decay over one sampling period $\Delta t$. The resulting maximum possible resolving power is easily calculated to be:\\

\begin{equation}
    \frac{\alpha }{\Delta \alpha } = \left( 1-e^{-\frac {\Delta t}{\tau _{qp}}}\right)^{-1}
\label{eq:tqp}
\end{equation}
In this example of a homodyne readout, sampling rates of about 1 MHz are common, equivalent to a $\Delta t$ of 1 $\mu s$ between data points. If we assume a quasi-particle recombination time of 100 $\mu s$ this sampling rate would limit the possibly achievable resolving power to R = 100.5. An assumed $\tau_{qp}$ of 25 $\mu s$ could limit R to 25.5. As the impossibility to distinguish between early and late arriving photons results in identical digital measurement values, no further noise reduction techniques, such as optimal filtering, can offer to improve these results - for more details, please see the supplementary materials. This effect can be significant for some superconductors but our assumption of a 1 MHz ADC sampling rate is only realistic for the discussed homodyne readout. 
\\
Standard readouts meant to be used with an MKID detector array and not just to prototype single pixels always use a heterodyne readout scheme: The incoming high-frequency signal (usually in the GHz range) is down-mixed to several hundred MHz, digitised and then undergoes Fast Fourier Transformation (FFT) in order to separate the individual channels/pixels. This means that two different sampling rates become relevant: The ADC sampling rate, which has to be much higher to allow to resolve all desired frequencies, and the much slower FFT sampling rate. As standard FFT algorithms \cite{udsp} do not snap-shot a varying input value but instead, for finite sampling times, result in an average over the sampling time, they are capable to distinguish between photons arriving just before or just after an FFT sample, and the above explained effect does not apply to the FFT sampling rate. It will still be relevant for the ADC sampling, but with typical ADC rates for heterodyne, full-array readouts of 4 GHz, the above discussed 100 $\mu s$ / 25 $\mu s$ quasi-particle recombination times only result in insignificant upper limits to R of $4\times 10^5$ and $1\times 10^5$  respectively.\\
One further note on this analysis is that the sampling mechanism described in this article is simplified and only represents a worst-case scenario for the degradation of the energy resolution due to the sampling frequency of the readout electronics.

\vspace{0.5cm}
There are 3 main sources of further noise / measurement inaccuracy in UVOIR MKIDs: Hot phonon loss to the substrate\cite{deVissersphononiccrystal}, an "excess" phase noise usually referred to as two-level systems (TLS) noise\cite{tls0} and Generation-Recombination (GR) noise. TLS noise is usually attributed to the coupling of electric dipole moments of contaminants or surface defects to the electric field of the resonator.  A semi-empirical model for TLS noise can be found, for example in \citet{tlsmodel1}. \\ Additionally, Generation-Recombination (GR) noise is produced by the dynamic generation and destruction of excess Cooper-pairs\cite{gr0} due to effects that are not induced by the optical detection. Their number in a superconductor is defined as a time-average as Cooper-pairs can form and break spontaneously. Given its nature, GR noise is material dependent and strongly depends on the critical temperature of the superconductor and the operation temperature of the MKIDs.\\A detailed discussion of TLS and GR noise goes beyond the scope of this publication, but to give a rough comparison: In our experimental setup we often probe MKIDs with tones at -90 dBm, and typically measure a phase noise of $\pm5^{\circ}$. This value for the baseline fluctuation without photons hitting the detector will only show noise contributions from HEMT, TLS and GR, and as discussed above, we expect the HEMT-amplifier phase noise to be around R = 91 equivalent to $\pm1.9^{\circ}$. We therefore attribute the typical remainder noise to TLS and GR in our devices. As soon as photons are detected, the random loss of phonons with energies above the superconducting bandgap to the substrate further widens the achieved spectral lines and thus reduces R. This hot phonon loss highly depends on pixel geometry details and has recently been clearly demonstrated by \citet{deVissersphononiccrystal}. Given the qualitative nature of our discussion on GR and TLS noise, we do not include their contributions to the limits on R in this paper.\\

If we want to combine the noise sources discussed above, we have to assume that they follow a Gaussian distribution and are independent. Under these assumptions, the width of the overall distribution is given by the root of the sum of the squares of the individual widths \cite{taylor-analysis}
\begin{equation}
    {\scriptstyle\frac{\Delta \alpha}{\alpha} = \sqrt{\left(\frac{\Delta \alpha}{\alpha}\right)_{HEMT}^2 + \left(\frac{\Delta \alpha}{\alpha} \right)_{Currents}^2+\left(\frac{\Delta \alpha}{\alpha}\right) _{Readout}^2+\left(\frac{\Delta E}{E}\right) _{Fano}^2}}
\label{eq:DAA_00249}
\end{equation}
We can combine expressions in $\alpha$ and E as in our simplified example the measured phase angle $\alpha$ is assumed to be proportional to the photon’s energy E. Following all our presented assumptions (2.1 K HEMT noise temperature, 0.8 K critical temperature, 100 µs quasi-particle lifetime, …) and with the values discussed above and repeated in Table \ref{tab:daa} we get a predicted upper limit for the achievable resolving power of R = 70.6 for a heterodyne readout or R = 57.8 for our homodyne example.\\

Most standard noise reduction procedures like e.g. optimal filtering  achieve improvements in R by analysing all points in the photon signal pulse instead of just its maximum. They are effective in reducing noise sources that add random variations to every measured data point like e.g. the HEMT amplifier noise or TLS and GR noise (see below). But the other noise sources discussed do not randomly vary single points in the photon pulse but result in variations of the overall pulse height instead. Current inhomogeneity, Fano variations and the effect of the ADC sampling rate vary the value of all points in the signal pulse in the same way and therefore can’t be filtered out easily.

\begin{table}[h!]
    \centering
    \begin{tabular}{|c|c|c|c|c|}
    \hline
     & $HEMT$ & $Currents$ & $Heterodyne\, Readout$ & $Homodyne\, Readout$ \\
     & & & $4GHz$ & $1MHz $\\
    \hline
    $\left(\frac{\Delta \alpha }{\alpha}\right)$ & 0.01 & 0.002 & $2.5\times10^{-6}$ &0.00995 \\
    \hline
        $R_{max}$  & 91.1 & 606.0 &$4\times10^5$ & 100.5 \\
    \hline
    \end{tabular}
    \caption{Contributions to the energy resolution and resolving power limits from the discussed sources of noise: HEMT-amplifier, Current inhomogeneity, sampling frequency.}
    \label{tab:daa}
\end{table}{}

Under our working assumptions (and ignoring further noise sources, please see above) the HEMT amplifier noise accounts for most of the limitations to R, followed by the Fano limit. Contributions due to current inhomogeneities are expected to be more than a factor of two smaller. For single-pixel, homodyne readout schemes R could see further limitations depending on quasi-particle lifetimes, but for heterodyne, full-array readouts this contribution is insignificant in most cases. This means if illuminated by a blue, $\lambda$ = 400 nm photon (our example wavelength), the theoretical MKID taken into consideration would be capable of achieving a maximum resolving power of R=70.6, equivalent to a $\pm0.04$ eV inaccuracy of the measured $3.01$ eV  photon energy. Contributions from further noise sources will of course further reduce the realistically achievable R. 

\begin{figure}[h!]
\includegraphics[scale=0.23]{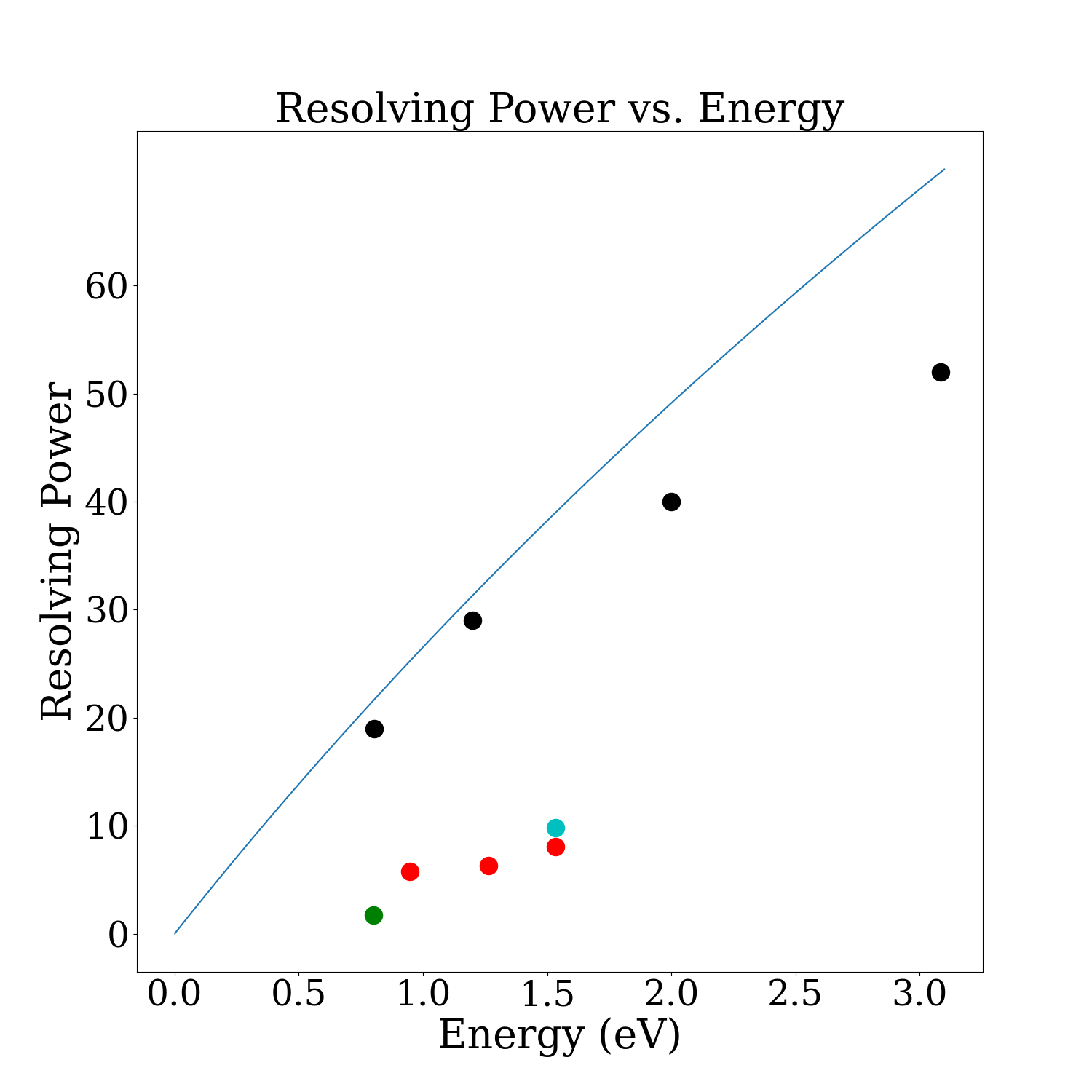}
\caption{Maximum resolving power $\frac{E}{\Delta E}$ as a function of the detected photon’s   energy. The coloured dots show measured values from: Green - \citet{Guo}, Cyan - \citet{mazin2020}, Red - \citet{Szypryt_2017}., Black - \citet{deVissersphononiccrystal}.}
  \label{fig:DAA-lambda} 
\end{figure}

\vspace{0.5cm}
The predicted maximum R = 70.6 of our example MKID assumes a $180^{\circ}$ phase shift for 400 nm photons. Less energetic photons will produce smaller signals, and $R_{max}$ will decrease with wavelength: The Fano limit decreases with photon energy (from 114.5 for 400 nm to 59.2 for 1500 nm in our example). The HEMT-amplifier noise is constant but as the absolute value of $\alpha$ changes with increasing wavelength $\lambda$, the signal to noise ratio and thus $R_{max}$ also decreases. The same is true for the contribution from current inhomogeneities. ADC-sampling related effects are wavelength independent. Based on these dependencies, we calculate a maximum R as a function of $\lambda$ using our assumptions and considering a resonator that would produce a $180^{\circ}$ phase signal for a 400 nm photon and $20^{\circ}$ for 1500 nm. Under these assumptions, Eq.\ref{eq:DAA_00249} becomes:
\begin{equation}
    {\scriptscriptstyle\frac{\Delta \alpha}{\alpha}\left(\lambda\right) = \sqrt{\left(\frac{\Delta \alpha}{\alpha}\left(\lambda\right)\right)_{HEMT}^2 + \left(\frac{\Delta \alpha}{\alpha}\left(\lambda\right)\right)_{Currents}^2+\left(\frac{\Delta \alpha}{\alpha}\right))_{Readout}^2+\left(\frac{\Delta E}{E}\right)\left(\lambda\right) _{Fano}^2}} 
\label{eq:DAA-lambda}
\end{equation}{}
The behaviour described by Equation \ref{eq:DAA-lambda} is shown in Figure \ref{fig:DAA-lambda}.  Also shown are data points for achieved resolving powers in literature, demonstrated  for MKIDs made of different superconductors: \citet{Guo},  (Green)  for their TiN/Ti/TiN MKIDs measured a value of  $R =3.7$ at $1550$ nm, \citet{Szypryt_2017}, (Red) reported PtSi MKIDs with a resolving power in the wavelength range $808- 1310$ nm between $8.1 - 5.8$; and \citet{mazin2020}, (Cyan) reports a resolving power of $10$ at $808$ nm for hafnium (Hf) MKIDs. \citet{deVissersphononiccrystal} (Black) reports resolving powers ranging between 19 and 52 for wavelength within $402-1545$ nm achieved by reducing hot phonon losses.
{}

\section*{Conclusions}

In this paper, we have modelled the  theoretical limits to resolving power of single-photon counting UVOIR Microwave Kinetic Inductance Detector; we have produced a model that accounts for HEMT-amplifier noise, current inhomogeneity and sampling frequency.  We describe the trend of the maximum achievable R varying with the wavelength of the incident radiation. Under our assumptions, amplifier noise is the second most limiting factor after TLS and Generation Recombination noise. We have also demonstrated that the Fano limit and current inhomogeneities contribute with smaller effects.

\section*{Supplementary Material}
An in-depth discussion of the handling of HEMT-amplifier noise and the current inhomogeneity induced noise, further discussion on the effects of digital sampling rates, as well as more details on the geometry of our used example MKID pixel can be found in the supplementary material.

\section*{Author Statements}
“Copyright (2024) Author(s). This article is distributed under a Creative Commons Attribution (CC BY) License.”
\section*{Author Contributions}
\noindent
\textbf{Mario De Lucia}: conceptualisation (supporting); methodology (lead); software (lead); writing – original draft (lead); Formal Analysis (lead).  \textbf{Gerhard Ulbricht} conceptualisation (lead); validation (lead); writing - review and editing (equal). \textbf{Eoin Baldwin} methodology (supporting). \textbf{Jack D. Piercy} validation (supporting); writing - review and editing (equal). \textbf{Oisin Creaner} validation (supporting); writing - review and editing (equal). \textbf{Colm Bracken} methodology (supporting); writing - review and editing (equal). \textbf{Tom P. Ray} writing - review and editing (equal); funding acquisition.
\section*{Acknowledgements}

The authors want to acknowledge P. J. de Visser for the many useful conversations on the matter.\\ 
This material is based upon works supported by the Science Foundation Ireland under Grant No. grant 15/IA/2880. 

\bibliography{aipsamp}

\end{document}